\documentclass[namedreferences]{kluwer}
\usepackage{graphicx}

\begin{document}
\runningtitle{REGULARIZED SOLAR FLARE HARD X-RAY SPECTRAL INDEX}

\runningauthor{E.P. KONTAR AND A.L. MACKINNON}
\begin{opening}

\title{REGULARIZED ENERGY-DEPENDENT SOLAR FLARE HARD X-RAY
SPECTRAL INDEX}

\subtitle{}

\author{Eduard P. \surname{KONTAR}\email{eduard@astro.gla.ac.uk}} \institute{Department of Physics
\& Astronomy, University of Glasgow, G12~8QQ, United Kingdom}

\author{Alexander L. \surname{MACKINNON}\email{alec@astro.gla.ac.uk}} \institute{Department of
Adult and Continuing Education, University of Glasgow, G3~6NH,
United Kingdom}

\begin{abstract}
The deduction from solar flare X-ray photon spectroscopic data of
the energy dependent model-independent spectral index is
considered as an inverse problem. Using the well developed
regularization approach we analyze the energy dependency of spectral
index for a high resolution energy spectrum provided by {\it
Ramaty High Energy Solar Spectroscopic Imager} (RHESSI). The
regularization technique produces much smoother derivatives while
avoiding additional errors typical of finite differences.
It is shown that observations imply a spectral index varying
significantly with energy, in a way that also varies with time as the
flare progresses. The implications of these findings
are discussed in the solar flare context.
\end{abstract}

\end{opening}

\section{Introduction}

Hard X-ray spectroscopy is considered to be an important tool for
the study of high energy processes at the Sun (e.g. Brown, 1971;
Lin and Hudson, 1976; Aschwanden, 2002), yielding vital, direct
information on fast electron populations in flares. Specifically,
the spatially integrated X-ray spectrum may be viewed as the
convolution in electron energy of the bremsstrahlung cross-section
and the mean electron flux in the source region (Brown, Emslie and
Kontar, 2003). The functional form of the photon energy spectrum
then contains information on the energy-dependence of the mean
electron flux. This information may in turn be interpreted to bear
on pictures of flare electron acceleration and propagation.

Even the earliest observations (e.g. Kane and Anderson, 1970)
revealed the overall power-law form of the photon spectrum
$I(\epsilon)$ (photons~cm$^{-2}$~s$^{-1}$~keV$^{-1}$) above about
10 - 20 keV: $I(\epsilon) \sim \epsilon^{-\gamma}$, for some
$\gamma > 0$. In the simplest possible (isotropic,
non-relativistic) treatments of bremsstrahlung production, a
power-law photon energy spectrum implies a source mean electron
flux $F(E)$ (electrons~s$^{-1}$) which is also a power-law in electron
energy $E$. Further, simple
assumptions about the post-acceleration propagation of electrons
in the source (e.g. thin target, in which all electrons escape
with negligible energy loss, or thick target, in which all
electrons stop completely in the source) then lead to $F(E) \sim
E^{-\delta}$, with simple, linear relationships in each of the
thick and thin target cases between $\delta$ and $\gamma$ (Brown,
1971).

A pure power-law form of $F(E)$ would be an
important clue to the nature of the acceleration process (Korchak,
1971). Power-law electron energy distributions are found commonly
in nature and are a natural consequence of certain acceleration
mechanisms (Miller et al, 1990; Petrosian et al., 1994). The
distinctive feature of a power law, however, is the absence of any
natural scale so it is probably true that any deviations from
power-law behaviour give more decisive clues to the acceleration
mechanism than the power law itself. Any such features would, of
course, be reflected in deviations from power-law behaviour of
$I(\epsilon)$ - see Lin and Schwartz (1987) for an example. In
particular, $F(E)$ of power-law form only up to some maximum
energy $E_{max}$ would result in an $I(\epsilon)$ which falls off
more and steeply as $\epsilon = E_{max}$ is approached.

Even for an accelerated pure power-law $F(E)$, electron transport
and/or radiation physics may produce deviations from power-law
behaviour in the observed spatially integrated photon spectrum.
Anisotropy of the mean electron distribution combined with the
directionality of bremsstrahlung emission for deka-keV electron
energies would also have this effect, possibly with diagnostic
potential for the source electron angular distribution (Massone et
al, 2004). Compton back-scattering of photons from the photosphere
(``X-ray albedo") will also distort a power-law photon spectrum
(Bai and Ramaty, 1978).

In a thick target, several physical processes may alter the
analysis that leads to a power-law $I(\epsilon)$ from a power-law
$F(E)$, e.g. nonuniform ionisation (Brown, 1973; Kontar et al,
2003). As emitting electron energies approach thermal energies,
the growing importance of velocity diffusion (as opposed to
systematic mean slowing-down) will cause the mean electron
distribution, and thus the photon spectrum, to deviate from a
power law (Galloway et al., 2004). Noncollisional energy losses
could have a wide variety of consequences for the photon spectrum
(Brown and MacKinnon, 1985).

It seems clear that deviations from power-law $I(\epsilon)$ hold
significant diagnostic potential. We may define $\gamma(\epsilon)=
-({\epsilon}/{I}){dI}/{d\epsilon}$ as an energy-dependent spectral
index, identically constant for a pure power-law $I(\epsilon)$,
but likely to be informatively non-constant much of the time. Most
earlier X-ray detectors had insufficient photon energy resolution
to fully explore this potential, however, but we may now
realistically aim to calculate $\gamma(\epsilon)$ numerically from
RHESSI (Lin et al, 2002) data.

With the simplest possible approximation (Kramers, 1923) to the
bremsstrahlung cross-section we see further the relationship
between the spectral index and the mean electron flux ${\bar F}
(E)$, in particular illustrating physically the connection between
derivatives and an inverse problem. The photon flux is a
convolution of the mean electron flux and the cross-section
(Brown, 1971). Making Kramers' approximation we can write
\begin{equation}\label{kramer}
I(\epsilon)=\frac{Q_0}{\epsilon}\int\limits_{\epsilon}^{\infty}\frac{\bar{F}(E)}{E}dE
\end{equation}
where $Q_0$ is a constant. From the observed flux $I(\epsilon)$ we
need to obtain the mean electron flux. Differentiating both parts
of (\ref{kramer}) immediately leads us to the following
\begin{equation}\label{kramer2}
\left.{\bar F} (E)\right|_{E=\epsilon}=\frac{\epsilon
I(\epsilon)}{Q_0}[\gamma (\epsilon)-1]
\end{equation}
Since the photon flux is given from observations, Equation
(\ref{kramer2}) shows that the spectral index uniquely determines
the mean electron flux, at least in the limit that Kramers'
approximation applies (Kramers, 1923).

Real data always come with noise, however, and numerical
differentiation of data is a noise-amplifying process. Thus there
is a need to establish reliable procedures for obtaining
$\gamma(\epsilon)$ (and more generally, ${dI}/{d\epsilon}$) safely
from data. Here we demonstrate that regularisation techniques
(Groetsch, 1984; Hanke and Scherzer, 2001) can be applied for our
purposes and yield such a robust procedure. The resulting smooth
function may then be safely differentiated to obtain a best
estimate of $\gamma(\epsilon)$. In Section 2 we give the formal
demonstration of such an approach, in the process clarifying the
sense in which this gives a ``best" estimate. Section 3 applies
this technique to obtain $\gamma(\epsilon)$ from RHESSI data for
the flare of 26 February, 2002 and study its time-dependence.
Section 4 gives conclusions and discusses physical implications of
the work carried out here.

\section{Derivative as an inverse problem}

Let us assume that we have a smooth function $y(x)$ over the
interval $x _{01}\leq x\leq x _{02}$. We have a finite sample
$y_{i}$ of measured values of this function, obtained over some
grid $x_{01}=x _{0}<x_{1}<...<x_{i}<...<x_{n}=x _{02}$ with mesh
size $\Delta x$. The noisy data set has an error
\begin{equation}\label{yerror}
|y_{i} - y(x_{i}) | \leq \delta y
\end{equation}
where $\delta y$ is an uncertainty of measurement.
%removed completely the bit about nonuniform errors and stuck if after end document - ALM, 30/8/04

We want to find the best smooth estimate of the derivative $y'(x)$
using the given data set $\forall$ $x \in [x_{01},x_{02}]$. The
two point finite difference estimate is readily available from the
Cauchy expansion with the following bound
\begin{equation}\label{yperror}
\left|\frac{y_{i+1}-y_{i}}{\Delta x}-y'(x_i)\right| \leq O(\Delta
x +\delta y/\Delta x),
\end{equation}
where the first and second terms in the right hand side represent
consistency and propagation errors respectively (Groetsch, 1984).
The first term in the right hand side comes from the discreteness
of the data set, while the second is connected with the errors of
the data $\delta y$.

Equation (\ref{yperror}) explicitly shows that for decreasing
$\Delta x$ the error in the estimated derivative deteriorates
rapidly, whereas for increasing $\Delta x$ the error grows only
linearly. The propagation error can be substantially suppressed by
taking larger $\Delta x$, a procedure also known as regularization
by coarse discretization. The extreme case of discretization
corresponds to a linear fit. Obviously, this is far from
desirable: although we have minimised the effects of error
propagation we have only a single estimate of $y'(x)$ across the
whole of our range of $x_i$. On the contrary, we want to extract
as much information as possible. The finite difference approach
also leads to a non-smooth estimate of $y'(x)$ - i.e. the
resulting derivative is piecewise continuous. Note that the
right-hand side of the Equation (\ref{yperror}) reaches its
minimum value which is $O(\sqrt{\delta y})$ when $\Delta x \sim
\sqrt{\delta y}$.

Let us now assume the existence of a function $f(x)$, which is a close
approximation to the data set:
\begin{equation}\label{c1}
\frac{1}{n-1}\sum _{i=1}^{n}(y_{i}-f(x_i))^2\leq (\delta y)^2,
\end{equation}
Below we describe a detailed procedure for the construction of
$f(x)$. We have to suplement this procedure with boundary
conditions, the values of $f(x)$ at two values of $x$. The values
$y_i$ at any two of the values $x_i$ would suffice in principle.
It might under some particular circumstances (e.g. suspected
outliers) be appropriate to choose two of the interior points, but
in what follows we simply choose the values at the endpoints, so
that boundary conditions become: $f(x_{01})=y_0$, and
$f(x_{02})=y_n$. Let us also assume that $f(x)$ is a square
integrable function over interval $x\in [x_{01},x_{02}]$, so we
can introduce a norm
\begin{equation}\label{norm}
||f||\equiv\left(\int_{x _{01}}^{x _{02}}
f(x)^2\mbox{d}x\right)^{1/2}
\end{equation}

Hanke and Scherzer (2001) describe an optimal approach to
constructing a smooth estimate of the derivative $y'(x)$.
Following them, we also require that the function $f(x)$ has
a smooth derivative, in the sense that
\begin{equation}\label{minf}
||f''||=\mbox{min}
\end{equation}
This means that we seek the function $f(x)$ whose second
derivative has the smallest norm. Roughly speaking, if we
represent $f(x)$ locally by the first two terms of its Taylor
expansion, $f''(x)$ gives an estimate of the error on $f'(x)$ and
minimising $||f''||$ minimises the error in the resulting estimate
for $f'$.

More specifically, let us estimate the error of our approximation
if our basic conditions (\ref{c1}),(\ref{minf}) are met. The
propagation error for the derivative $f'$ can be estimated by
considering the following norm
\begin{equation}\label{fnorms}
||f'-y'||^2\leq||f-y||\;\;||f''-y''||
\end{equation}
where we integrated by parts and used the Cauchy-Schwarz
inequality. The first factor in the right hand side is bounded by
our requirement (\ref{c1}). The second factor is also limited as
soon as $||f''||\leq || y''||$, which is true as long as equation
(\ref{minf}) is satisfied. Now using the Minkowski inequality we
arrive at the following result
\begin{equation}\label{ferror}
|f'(x_i)-y'(x_i)|\leq O(\Delta x +\sqrt{\delta y})
\end{equation}
where we added consistency error in the same manner as in
(\ref{yperror}).

Comparing (\ref{ferror}) and (\ref{yperror}) we see the drastic
difference. The former shows no growth for small values of $\Delta
x$: fitting a suitably smooth function prior to estimating the
derivative eliminates the uncertainty associated with estimates
based on discrete bins. Moreover, the error (\ref{ferror}) is
bounded by the minimum in (\ref{yperror}).

The result of this formal exercise can be formulated in terms of
optimization (Hanke and Scherzer, 1998). Collecting our two
requirements (\ref{c1}),(\ref{minf}) into a single equation, we
require to minimise the functional:
\begin{equation}\label{func}
\Phi (f)\equiv \frac{1}{n-1}\sum _{i=1}^{n-1}(y_{i}-\int
_{x_{0}}^{x_{i}}f'(\xi)\mbox{d}\xi-y_0)^2 + \lambda ||f''(x)||^2
\end{equation}
among all smooth functions $f$ with $f(\epsilon _{01})=y(x
_{01})$, $f(x _{02})=y(x_{02})$, where $\lambda$ is so that
\begin{equation}\label{tikhonov}
\frac{1}{n-1}\sum _{i=1}^{n}(y_{i}-f_{\lambda}(x_i))^2 = (\delta
y)^2
\end{equation}
where $f_{\lambda}(x_i)$ is a solution of minimum problem
(\ref{func}). This is formally equivalent to an inverse problem
and the solution method employed is well-known as Tikhonov regularization
(Tikhonov, 1963). The equation (\ref{tikhonov}) is also known as a
discrepancy principle. It is interesting to compare the problem
(\ref{func}) with the inversion of the photon data using second
order regularization (see Kontar et al, 2004). The two problems
are virtually identical. The only difference is the operator: the
bremsstrahlung cross-section used in case of inversion of photon
data, while our operator is an integrator. For the numerical
implementation of this exercise, it is trivial to write
the formal solution of the problem (\ref{func}) using Generalized
Singular Value Decomposition of our two operators: integrator and
second order derivative (Kontar et al, 2004).

\begin{figure}
\label{exfig}
\begin{center}
\includegraphics[width=80mm]{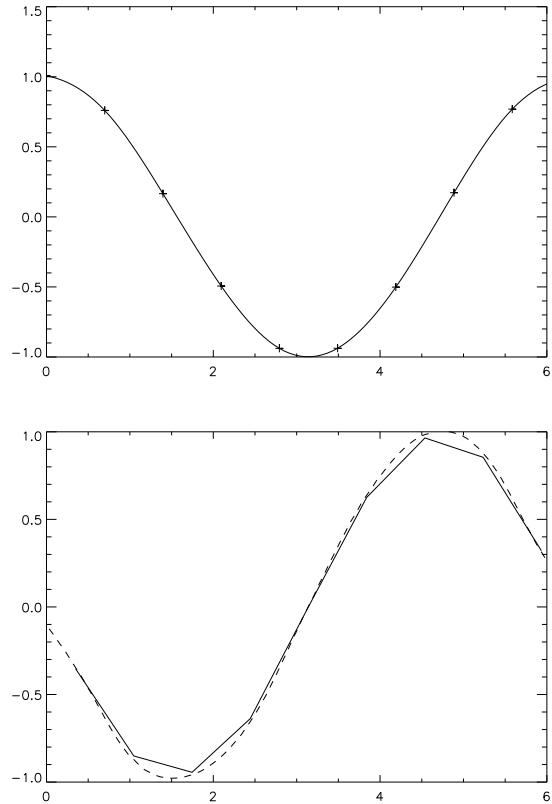}
\end{center}
\caption{Numerical example. Function $I(x)=\cos(x)$ (solid line)
and data points (crosses) (upper panel) and its derivative (lower
panel). Solid line is a regularized derivative for data points.
The dash line shows the actual values of finite differences.}
\label{example}
\end{figure}

The solution of the problem (\ref{func},\ref{tikhonov}) presents a
regularized solution of the problem for $f(x)$.
Figure~\ref{example} shows an example of this process applied to
the function $y(x)=\cos(x)$, with a perturbation added in the form
of random noise at the level of $1\%$ of the value of $y$. The
upper panel shows the actual data points generated in this way,
together with the original function to emphasise how close the
``data" points generated actually are. The lower panel shows
values of the derivative calculated just by simple differencing of
the data points, and the smooth curve generated by first
regularising the data in the way described above. Judging from
this, the derivative resulting from the regularization procedure
is clearly much better. Compared to this, simple differencing of
data with even this modest level of noise destructively affects
the result.

\section{Energy variation of spectral index in a solar flare}

We defined the energy-dependent spectral index as a
logarithmic derivative: $\gamma(\epsilon)\equiv -\mbox{d}\ln
I/\mbox{d}\ln \epsilon$, where $I(\epsilon)$ is an observed
spectrum given as a data set $I_i$ for every $\epsilon _i$.
Choosing quantities $x_i$ and $y_i$ as follows
\begin{equation}\label{int}
\ln(I_i)\rightarrow y_i , \;\; \ln(\epsilon_i)\rightarrow x_i
\end{equation}
we find our desired spectral index to be $\gamma
(\epsilon)=-f'(x)$, where $f(x)$ is determined from the measured
values $y_i$ as described in the previous section. We note here
that since the bremsstrahlung photon spectrum is a convolution of
cross-section and electron flux (Brown, Emslie and Kontar, 2003),
logarithmic derivative of photon spectrum should always exist.

\begin{figure}
\begin{center}
\includegraphics[width=120mm]{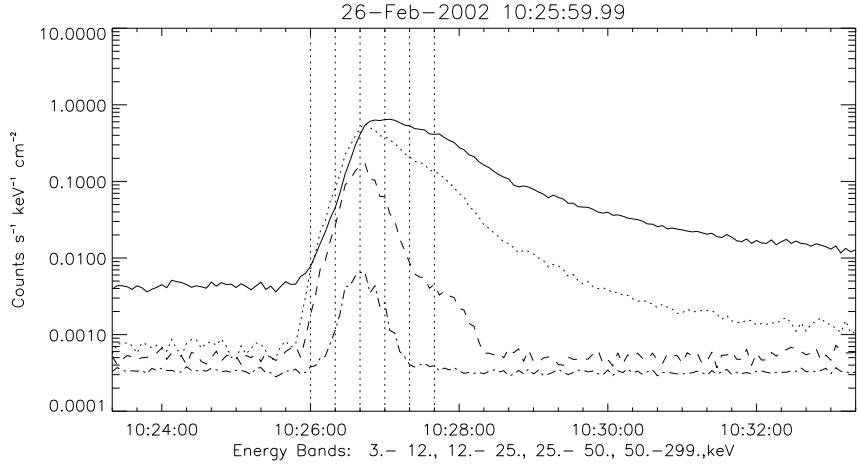}
\end{center}
\caption{Temporal variation (4 seconds cadence) of the count rates
in seven front RHESSI segments for the February 26, 2002 solar
flare. The vertical lines show five 20 second selected
accumulation intervals for spectral analysis.} \label{flux}
\end{figure}

We now give a first, illustrative application using the RHESSI
data from the GOES M-class flare that happened on February 26,
2002 around~10:27~UT (Figure \ref{flux}). This event provides good
photon count statistics while below the level of nonlinear
features of the instrumental response such as pulse pile-up in the
detectors (Smith et al, 2002). (These features are not fully
understood and their details lie outside the scope of this paper.)
Using standard software tools in SPEX we extracted the photon
spectral flux (photons~keV$^{-1}$~s$^{-1}$~cm$^{-2}$) in the range
above $10$keV accumulated in seven front segments out of nine
detectors. We omitted detectors 2 and 7 due to low operational
energy resolution at the moment of the observation.

\begin{figure}
\begin{center}
\includegraphics[width=80mm]{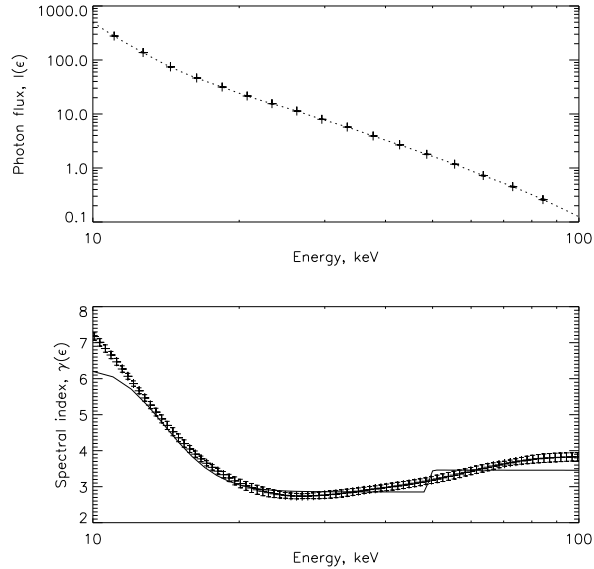}
\end{center}
\caption{Spectrum of February 26, 2002 solar flare for
10:26:40-10:27:00~UT (upper panel crosses). The low panel shows
the inferred spectral index with errors. The thick solid line
presents the spectral index based on the isothermal plus broken
power-law fit with $\gamma _{low}=2.8$,$\gamma_{upper}=3.5$ with
the break energy 49~keV.} \label{spectr}
\end{figure}

Figure \ref{spectr} shows the spectrum of the solar flare observed
by RHESSI on February 26, 2002 for the time interval
10:26:40-10:27:00~UT near at the peak of the flare. The lower
panel shows the spectral index, $\gamma(\epsilon)$, calculated via
the regularisation process described above. To construct error
bars we first generated two further sets of ``data", one $1
\sigma$ above and one $1 \sigma$ below the original data points
$I(\epsilon_i)$, where $\sigma$ is the uncertainty on the photon
spectrum. Then we applied the same regularisation procedure to
these two new data sets. The resulting confidence interval in
$\gamma (\epsilon)$ allows us to highlight several features of
interest, particularly in comparison with a other parametric
descriptions of the spectrum and by implication its derivatives,
for example an isothermal plus broken power-law spectral fit (Fig.
\ref{spectr}) - cf. Holman et al. (2003).

Most immediately obvious (Fig. \ref{spectr}) is the nonconstancy
of the spectral index within the energy range studied (10-100keV).
At the lowest photon energies the regularized spectral index has a
higher value than the index suggested by the isothermal plus
power-law fit. This behaviour of spectral index may indicate X-ray
emission from plasma with a range of temperatures. Similar
suggestions of non-isothermality have been made by Piana et al.
(2003), analyzing the mean electron flux spectrum deduced in the
July 23 2002 flare and comparing to an isothermal and broken
power-law fit (Holman et al, 2003). The regularized spectral index
$\gamma(\epsilon)$ also departs significantly from the broken
power-law fit near the break energy $49$ keV and above $70$ keV.
At the highest energies studied, $\gamma (\epsilon)$ approaches 4,
larger than the value found assuming a broken power-law. Above the
thermal component, the spectral index grows approximately linearly
with energy.

\begin{figure}
\begin{center}
\includegraphics[width=120mm]{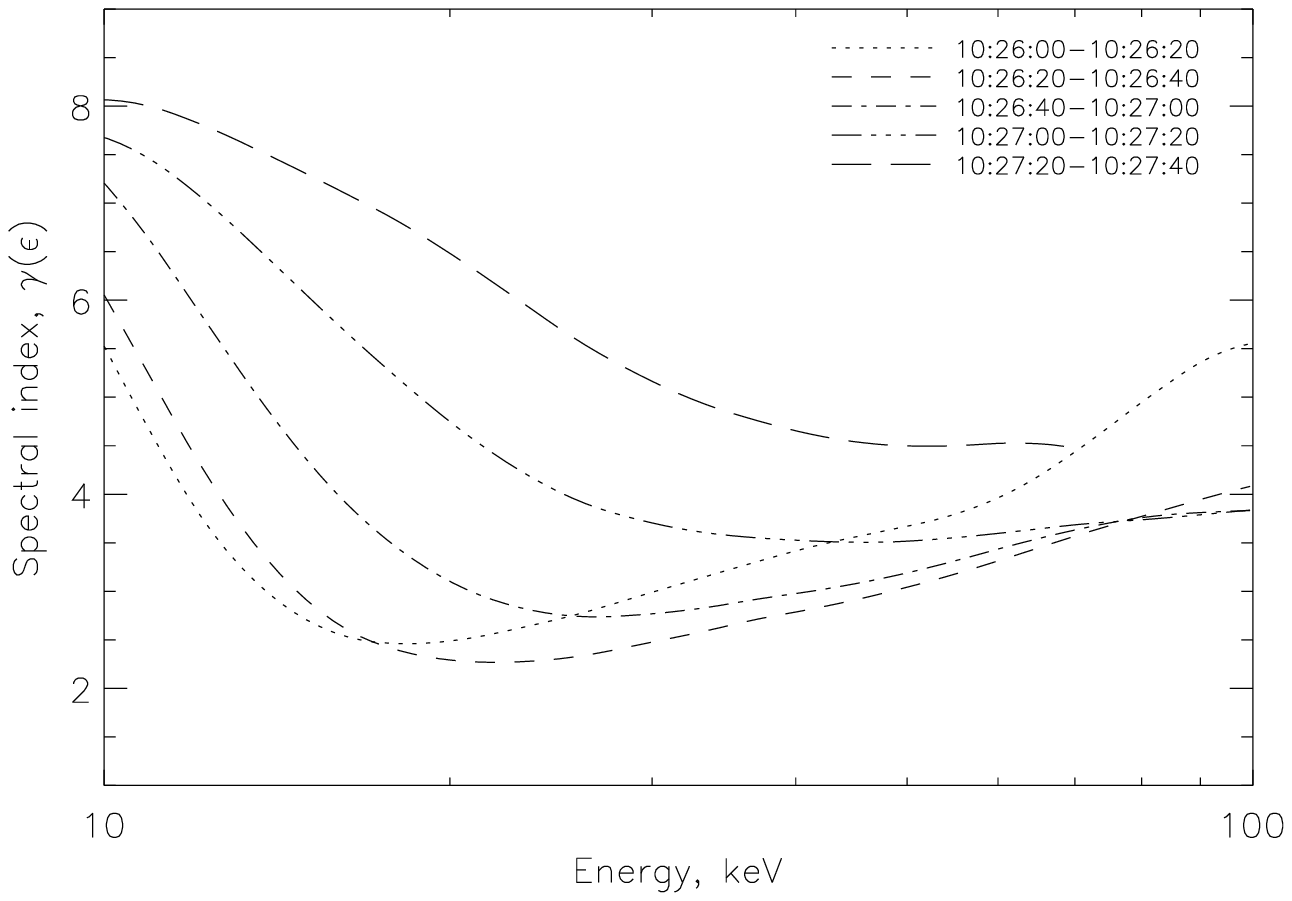}
\end{center}
\caption{Temporal variation of energy dependent spectral index.
Each line corresponds to one time interval.} \label{g_time}
\end{figure}

Figure \ref{g_time} shows that for all time intervals the
regularized spectral index shows a minimum, the position of which
grows with time. There is a clear minimum for the first time
interval at 10:26:00~UT near $17$ keV, while at 10:27:20~UT the
minimum is less clear and as high as $50$ keV. Additionally, the
spectral index varies less with energy as the flare progresses.
The spectral index growths with energy by as much as ~3.0 for the
first time interval, but it varies only by ~0.1 for the last
interval.

Earlier data from scintillation detectors established a
relationship between spectral hardness (parametrised via a single
value of $\gamma$ derived from fitting across the available energy
range) and total photon flux above some energy (Kane and Anderson,
1970). Specifically, $\gamma$ obtained in this way is
anti-correlated with total photon flux, so that the flare X-rays,
viewed crudely, have the hardest spectrum when they are most
intense. This ``soft-hard-soft" spectral behaviour has been
confirmed with RHESSI using ratios of photon flux in fairly broad
bands (Fletcher and Hudson, 2002; Hudson and Farnik, 2002).
Figure~\ref{g_time} shows the full, complex behaviour behind this
simplified description. Values of $\gamma(\epsilon)$ are indeed
lowest, for any given $\epsilon$, around the peak of the event,
but the decrease before and increase after this time do not occur
at the same rate for all $\epsilon$. The flare starts with
spectral index strongly dependent on energy, while a single
power-law fit produces a mean spectral index which is a
combination of small spectral index around $19$ keV and relatively
large $\gamma (\epsilon)$ at $70$ keV. In fact the minimum of the
spectral index shows a tendency to grow as the flare progresses.
Part of the transition from soft to hard spectra can thus be
understood in terms of a reduction in the degree of variation with
energy of $\gamma (\epsilon)$ in the range $20-100$ keV.

\begin{figure}
\begin{center}
\includegraphics[width=80mm]{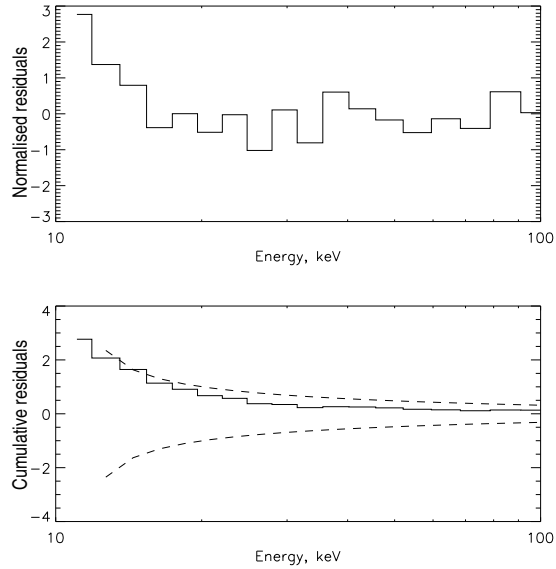}
\end{center}
\caption{The residuals and the accumulated residuals for the time
interval 10:26:40-10:27:00~UT (Fig. \ref{spectr}). The dash lines
in the low panel present $3\sigma$ level assuming residuals are
statistically independent.} \label{g_err}
\end{figure}

\section{Discussion}

The illustration above shows vividly how far real spectra depart
from the idealised power-law form. In this section we discuss some
possible physical implications.

In standard models of hard X-ray production (e.g. Aschwanden,
2002) the X-ray emitting region is located well above the dense
photosphere. X-rays emitted downwards will scatter on electrons of
the dense layers of the solar atmosphere, finding their way back
to an observer. The scattered flux, a function of the flare
location and primary spectrum, may produce a detectable alteration
of the net spectrum in the range 15-60 keV (Bai and Ramaty, 1978).
Assuming the primary photon spectrum to be a simple power law (Bai
and Ramaty, 1978) the photospheric albedo will peak at energies in
the 20 - 40 keV range, with the reflected flux $\approx 30-70\%$
of the primary flux. As a result, even if the primary spectrum is
a pure power-law, the observed spectrum is flatter before the
albedo peak energy and steeper above. This should be seen in the
energy variation of the spectral index. For the February 26, 2002
flare, at heliocentric angle $\approx 75^o$) and with photon
spectral index $\approx 3$, the results of Bai and Ramaty (1978)
suggest that the albedo contribution to the observed spectrum will
maximise near 35 keV, implying a lower spectral index below this
energy and higher above. This tendency can indeed be seen in
Figure (\ref{spectr}). Nevertheless, albedo alone cannot account
for the temporal variations shown in Figure~\ref{g_time}, however
- these must reflect properties of the primary X-ray spectrum, and
thus of the emitting electron population.

Suppose that there is a high energy cut-off (or substantial
softening) in the bremsstrahlung-emitting electron spectrum, at
some energy $E_{max}$. For energies close to $E_{max}$ the
spectral index will be a growing function of $\epsilon$. Since the
photon flux at a given energy is an integral over all electrons
above this energy, the change of the spectral index due to a high
energy cut-off can be seen at photon energies $\epsilon$ well
below $E_{max}$ (Kontar et al, 2004). During the first time
interval the variation with $\epsilon$ of the spectral index is
most significant, while the photon flux above 50~keV is very low
(Figure \ref{flux}). These facts together suggest a small number
of electrons above 50~keV at this early time, while later in the
flare the number of high energy electrons (above 100~keV) grows
substantially. This is at the very least a plausible
interpretation of the findings of Figure~\ref{g_time}, suggesting
that the ``soft-hard-soft" pattern of spectral evolution really
reflects changes in the highest electron energies present at
different times.

A number of other physical processes may produce detectable
variation in spectral index, even given an injected power-law
electron energy distribution. For example, it was suggested
(Brown, 1973; Kontar, Brown, and McArthur, 2002) that observed
deviations from power-law can be interpreted as a manifestation of
nonuniform target ionization. However, the possible variations of
spectral index are unlikely to exceed $0.6$ (Brown, Emslie, and
Kontar, 2003). Moreover, the minimum of the spectral index if
associated with chromospheric depth should show modest growth
after the impulsive phase of the flare (Kontar et al, 2003), while
the minimum of the spectral index shows no correlation with X-ray
flux.

%moved the spectral index/meanspectrum bit to the intro

\section{Conclusions}

We have shown how best to estimate numerically the energy-dependent X-ray
spectral index $\gamma(\epsilon)$ from a set of observed photon
fluxes at discrete energies, in a way that avoids the noise
amplification of numerical differentiation. Formally the necessary
procedure is equivalent to regularization. In consequence we have
shown that the spectral index of at least one solar flare can have
large variations with energy. The spectral index shows a clear
minimum in the range 17-50 keV and the value of this minimum
decreases as the flare progresses. The spectral index also tends
towards energy-independence as the flare progresses, suggesting
that something more like an ideal power-law photon spectrum is
eventually attained. The origin of such variation can be connected
with variability in the highest energy of the X-ray producing
electrons. For later times the high energy cut-off is higher,
thus producing a more uniform spectral index.
This seems a simple, plausible explanation of the spectral
behaviour found here; in particular it serves to illustrate the
value of obtaining and studying $\gamma(\epsilon)$.

\subsection*{Acknowledgment}
The authors are thankful to John Brown, Brian Dennis, and Hugh
Hudson for support, inspiring discussions and useful comments on
the manuscript. This work is supported by a PPARC Rolling Grant.

\end{document}